# FLUCTUATING SUNSPOT NUMBERS EXHIBIT A NON-MARKOVIAN DAMPED STOCHASTIC PROCESS

**Reynan L. Toledo[a], Reinabelle Reyes[b], and Christopher C. Bernido[a,c,1]**

[a] *Physics Department, University of San Carlos, Talamban, Cebu City 6000, Philippines*
[b] *National Institute of Physics, University of the Philippines, Diliman, Quezon City 1101, Philippines*
[c] *Research Center for Theoretical Physics, Central Visayan Institute Foundation, Jagna, Bohol 6308, Philippines*

## ABSTRACT

The rise and fall in the number of sunspots have served as a lynchpin in many investigations on solar dynamics. Arising from magnetic disturbances in the sun, variations in sunspot numbers have helped define a solar cycle of around eleven years which to date is yet to be fully understood. We model the fluctuation of sunspot numbers as a modulated Brownian motion characterized by a memory parameter $\mu$ and a decay parameter $\beta$. By matching the theoretical and empirical mean square deviation of the sunspot numbers, the values of $\mu$ and $\beta$ are determined for each solar cycle. This allows us to obtain an exact form of a probability density function (PDF) which closely matches the dataset for sunspots. This novel PDF for sunspot numbers exhibit a memory behavior from which some insights could be obtained. In particular, the values of $\mu$ indicate that consecutive sunspot numbers are negatively correlated for large times. The values of $\beta$, on the other hand, when viewed as a time series from one solar cycle to another, indicate a positive trend towards increasing values which could possibly suggest a diminishing solar activity.

## 1. INTRODUCTION

Solar activities and disturbances in the sun's magnetic field can damage the earth's telecommunication system, power generation, and can cause havoc to our satellites and space exploration. Interplanetary disturbances caused by solar eruptions can have severe effects upon reaching the earth (Liu et al 2014; Richardson and Cane 2010). In understanding solar dynamics, a prominent feature has been the solar cycle that spans around 11 years. Solar cycles are characterized by the flipping of the sun's magnetic north and south poles that exchange their positions accompanied by the rise and fall of the sunspot numbers. Variations in the sun's magnetic field have been associated with fluctuations in the number of observed sunspots (Okamoto and Sakurai 2018; Goelzer et al 2013). Among the many experimental and theoretical investigations on solar activities (Smith and Balogh 2008; Arlt et al 2013; Goelzer et al 2013; Hathaway 2015), some attribute a memory to solar cycle dynamics (Aschwanden & Johnson 2021; Karak & Nandy, 2012; Muñoz-Jaramillo et al, 2013;

---
[1]Corresponding author: cbernido.cvif@gmail.com

Yeates et al 2008). The paper of Muñoz-Jaramillo et al (2013) has shown that memory is limited to only one solar cycle. On the other hand, studies on solar flare events and flare clustering by Aschwanden and Johnson (2021) demonstrated that the sun reveals memory from a few hours to several decades.

We investigate the memory behavior of solar cycles in this paper by characterizing the fluctuations in sunspot numbers as a non-Markovian stochastic process. In particular, the fluctuations in sunspot numbers are parametrized in terms of a modulated Brownian motion with a memory parameter $\mu$ and a decay parameter $\beta$. We evaluate the mean square deviation (MSD) of the number of sunspots per solar cycle and match this with the MSD of the stochastic model. This allows us to obtain a probability density function corresponding to the dataset for each solar cycle. Comparison of the parameters $\mu$ and $\beta$ from one cycle to another provides insights into the memory behavior of the solar cycles considered. The stochastic framework that we use has earlier been successfully applied to other complex systems with memory such as the gelation of fibrin (Aure et al. 2019), the Great Barrier Reef degradation (Elnar et al 2021), nucleotide sequences in a genome (Violanda et al 2019), the diffusion of proteins with varying length (Barredo et al 2018), timing residuals of pulsars (Reyes & Bernido 2023), and cyclone tracks (Bernido et al 2015), among others.

## 2 Data and Methods

### 2.1 Sunspot Numbers

Data for the daily solar sunspot numbers (SSN) from 1818 to the present (Fig. 1) was obtained from the World Data Center SILSO, Royal Observatory of Belgium, Brussels. We investigate each of the eighteen solar cycles (SC) from SC 7 to SC 24.

Although solar cycles 7, 8, and 9 were considered, these cycles have a lot of missing data points. In particular, the percentage of missing values are 22%, 38%, and 13% for SC 7, SC 8, and SC 9, respectively. This is exemplified in Fig. 2a for SC 7. For these, the missing values in between two non-missing values were interpolated by the slope of a straight-line equation for the two non-missing values. The result is shown in Fig. 2b for SC 7. This is done only for solar cycles 7, 8, and 9.

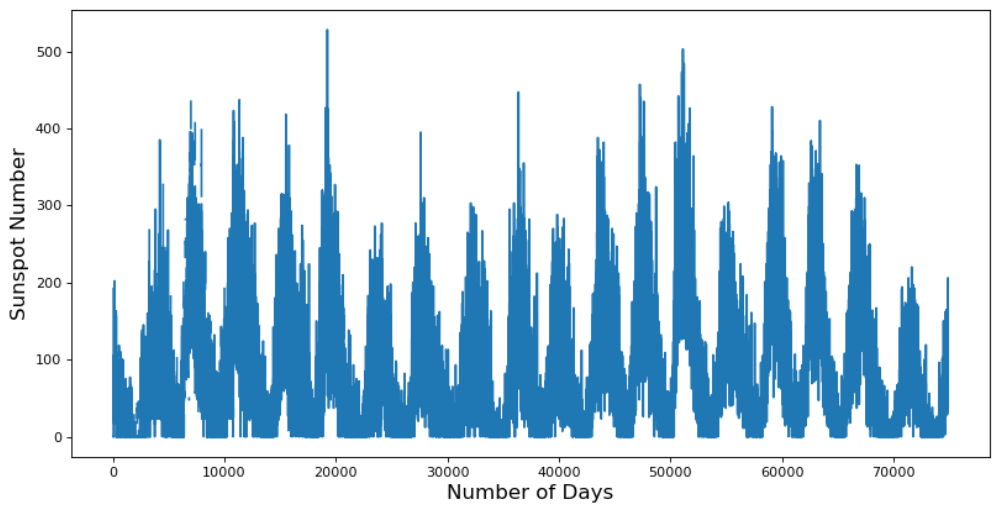

**Figure 1.** Daily sunspot numbers from 1818 to the present. One minimum point to another represents one solar cycle.

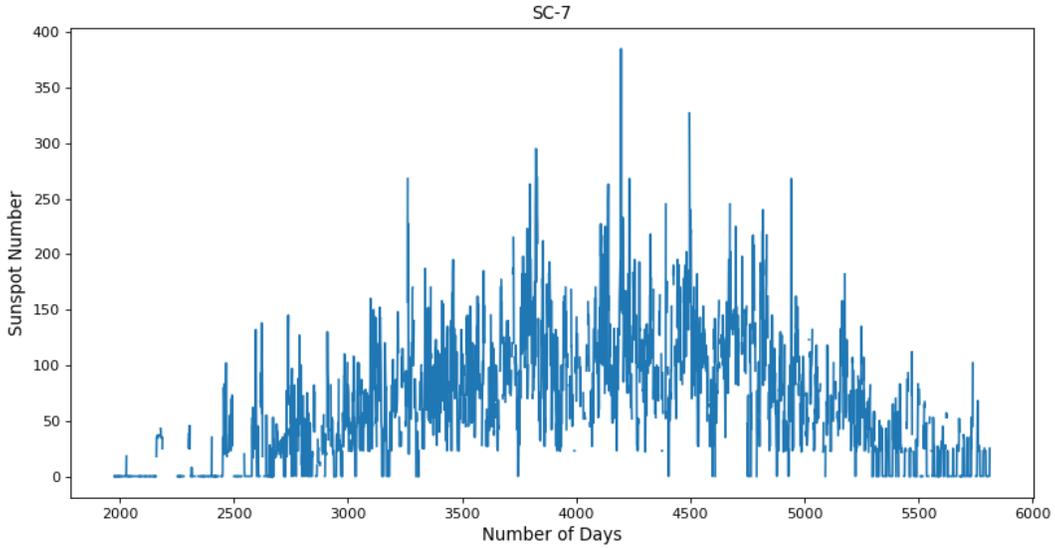

(a)

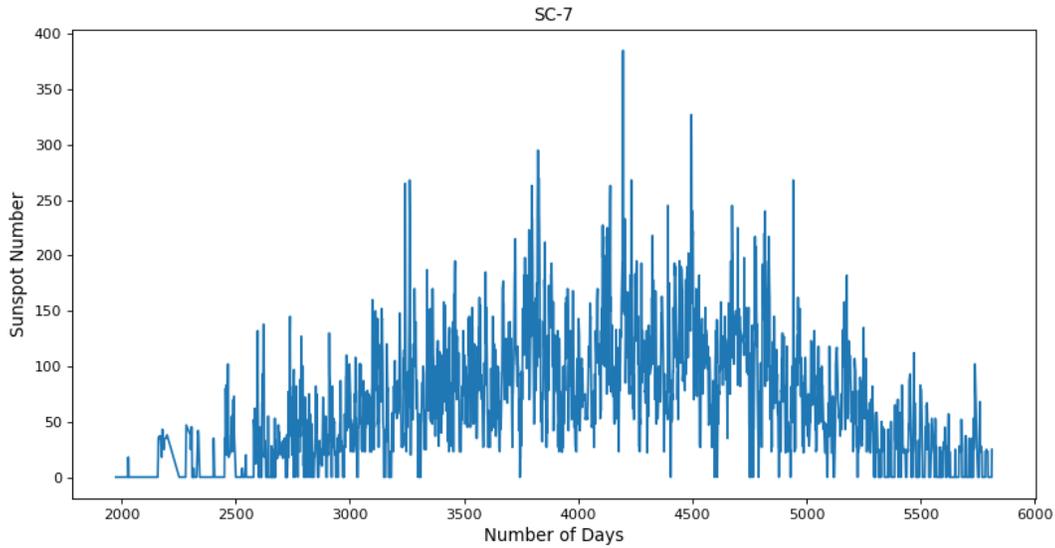

(b)

**Figure 2.** (a) Sunspot numbers for SC 7 with missing values; (b) SC 7 with missing values interpolated by the slope of a straight-line equation of two non-missing values.

The best fit for the sunspot numbers for each cycle (see, e.g., Fig. 3) has earlier been studied by Hathaway (2015). To investigate the stochastic nature of sunspot numbers, we focus on the fluctuations or deviations from the best fit given by the Hathaway function. Hence, the difference between the sunspot number and the Hathaway function at a given time is obtained. The data we will work on, which we simply refer to as sunspot numbers in our discussion, are the results of subtracting the original data values from the fitted Hathaway function as illustrated in Fig. 4 for SC 24.

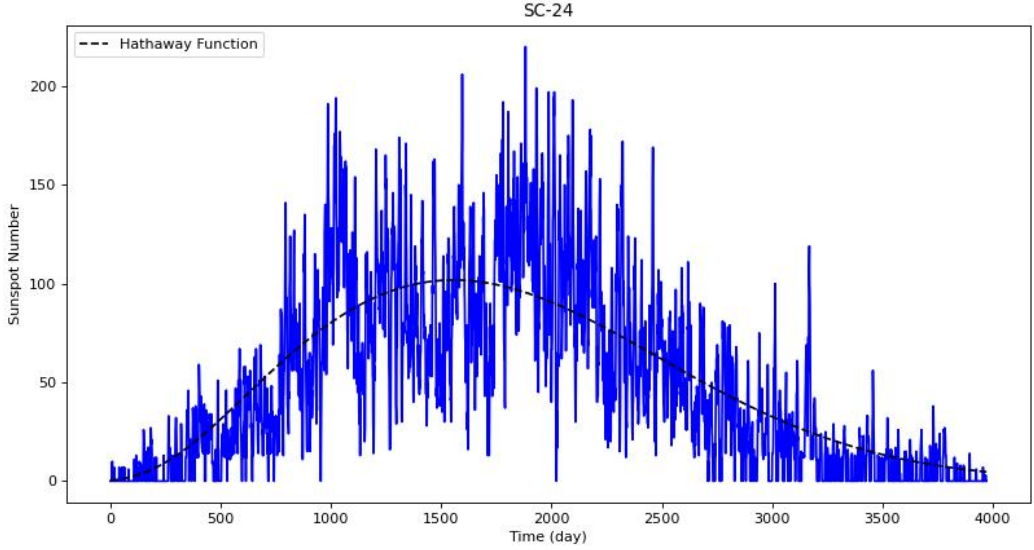

**Figure 3.** Sunspot numbers for SC 24 with the fitted Hathaway function.

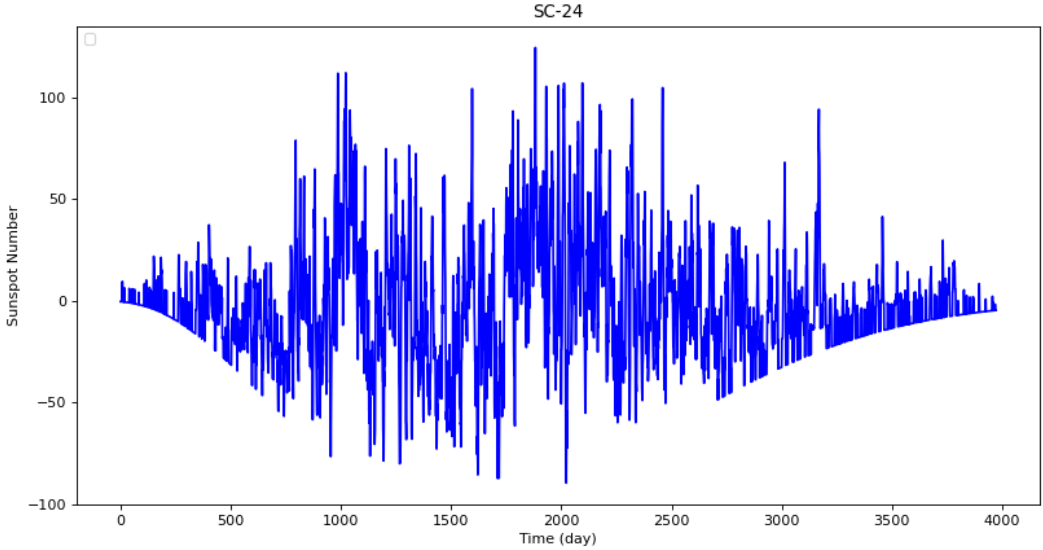

**Figure 4**. Fluctuating sunspot numbers obtained by subtracting the Hathaway function from the original data of SC 24.

## 2.2 Damped Stochastic Process with Memory

As a stochastic framework for investigating the rise and fall of sunspot numbers, we represent the fluctuating sunspot numbers, such as those in Fig. 4, by $x(t)$ where $t$ is time. We then parametrize the fluctuations in terms of a modulated Brownian motion $B(t)$, with the range of $t$ given by, $0 \leq$ $t \leq T$. Specifically, we write, $x(T) = x_i +$ $fluctuations$, as

$$x(T) = x_i + \sqrt{a} \int_0^T (T - t)^{\frac{(\mu-1)}{2}}$$

$$\times \, e^{\frac{-\beta}{2t}} \, t^{\frac{-(\mu+1)}{2}} dB(t) \,, \tag{1}$$

with, $x(0) = x_i$, as the initial value. In Eq. (1), the factor, $(T - t)^{(\mu-1)/2}$ acts as a memory kernel describing the physical system and, $e^{-\beta/2t}\,t^{-(\mu+1)/2}$, dampens the Brownian motion $B(t)$. The $\mu$ and $\beta$ are the memory and decay parameters, respectively. The parameter $a$ in Eq. (1) gives the amplitude of the fluctuation. Here, we set, $a = 80(maxSSN)$, as discussed in Section 3.1 where $maxSSN$ is the 13-month mean maximum value of each cycle obtained from the article of Hathaway (2015). The parameters $\beta$ and $\mu$ are determined from the dataset for each solar cycle.

The probability density function (PDF) corresponding to Eq. (1) is given by (see Appendix A),

$$P(x_f, T; x_i, 0) =$$
$$\frac{1}{\sqrt{2\pi a\,\Gamma(\mu)\beta^{-\mu}T^{\mu-1}e^{-\beta/T}}}\exp\left[\frac{-(x_f-x_i)^2}{2a\,\Gamma(\mu)\beta^{-\mu}T^{\mu-1}e^{-\beta/T}}\right]$$
$$(2)$$

where $\Gamma(\mu)$ is the Gamma function, and $x_f$ is the value of the fluctuating variable at final time $T$, i.e., $x(T) = x_f$.

### 2.2.1 Theoretical Mean Square Displacement

The mean square displacement (MSD) of the fluctuating variable, Eq. (1), can be evaluated as, $\text{MSD} = \langle x^2 \rangle - \langle x \rangle^2$, where $\langle x^2 \rangle = \int_{-\infty}^{+\infty} x^2\,P(x,T;x_i,0)\,dx$, with $P(x,T;x_i,0)$ given by Eq. (2). This yields (Bernido and Carpio-Bernido, 2015),

$$\text{MSD} = a\,\Gamma(\mu)\beta^{-\mu}T^{\mu-1}e^{-\beta/T}. \qquad (3)$$

The parameters $\beta$ and $\mu$ are determined by matching the theoretical MSD, Eq. (3), with the MSD of the fluctuating sunspot numbers from the dataset.

### 2.2.2 Empirical Mean Square Displacement

Consider a SSN dataset with $N$ datapoints in a solar cycle. We designate $x_j$ as the number of sunspots at time $t_j$ where, $j = 0,1,2,\dots,N$. Hence, the number of sunspots at the initial time $t_0$ is given by $x_0 = x(t_0)$, and the last sunspot number at time $t_N$ is given by $x_N = x(t_N)$ where $t_N$ corresponds to our final time $T$ in Eqs. (1) to (3). The empirical MSD is evaluated using,

$$\text{MSD}(\Delta) = \frac{1}{N-\Delta+1}\sum_{j=0}^{N-\Delta}\left(x_{j+\Delta} - x_j\right)^2, \quad (4)$$

with $N$ as the total number of datapoints in a solar cycle and, $\Delta < N$, as the lag time. The lag time $\Delta$ is the interval or separation between any two sunspot numbers. An increasing $\Delta$ generates the plot for the empirical MSD which can then be compared with the theoretical MSD given by Eq. (3).

## 3. Results and Discussion

### 3.1 Empirical and Theoretical MSD

Matching the theoretical and empirical MSD's showed values of parameter $a$ which appear to be correlated with the maximum value of the SSN (see, Fig. 5). To minimize the number of parameters we set, $a = 80(maxSSN)$, leaving us with just the memory parameter $\mu$ and decay parameter $\beta$ to be determined from the dataset.

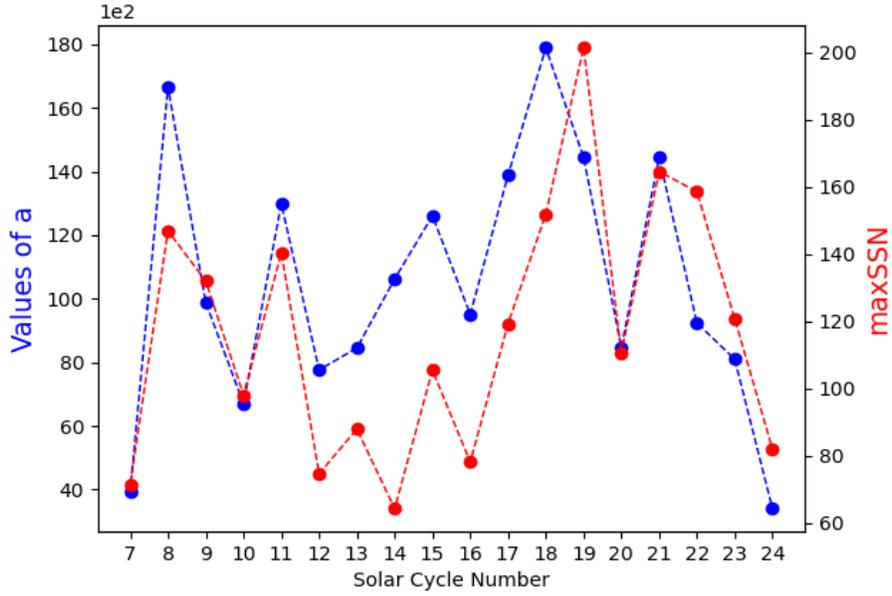

**Figure 5**. Values of parameter $a$ compared with the maximum value of SSN for different solar cycles.

The comparison between the theoretical MSD, Eq. (3), and the MSD of the sunspot numbers obtained using Eq. (4) with, $a = 80(maxSSN)$, determines the values of parameters $\beta$ and $\mu$ for each solar cycle. As shown in Fig. 6, the theoretical MSD captures the general shape of the empirical MSD for solar cycle 13. Similar results are obtained for the other solar cycles (see, Appendix B). The values of parameters $\beta$ and $\mu$ for each solar cycle are summarized in Table 1.

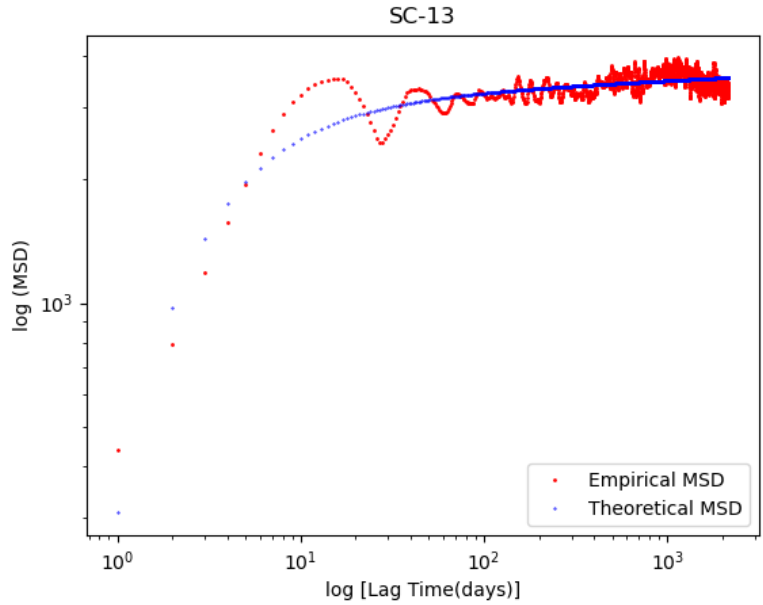

**Fig. 6.** The theoretical MSD, Eq. (3), captures the general shape of the MSD obtained from the sunspot numbers for solar cycle 13.

**Table 1.** Values of the parameters used in Eq. (3) for each solar cycle.

| Solar Cycle Number | $\beta$ | $\mu$ |
|---|---|---|
| 7 | 2.076931 ± 0.007410 | 1.048445 ± 0.000766 |
| 8 | 2.461346 ± 0.011974 | 1.056226 ± 0.001098 |
| 9 | 2.245219 ± 0.010902 | 1.046915 ± 0.001009 |
| 10 | 1.884186 ± 0.009071 | 1.032492 ± 0.001005 |
| 11 | 2.334695 ± 0.007919 | 1.019628 ± 0.000729 |
| 12 | 2.328903 ± 0.009691 | 1.031069 ± 0.000900 |
| 13 | 2.274251 ± 0.007969 | 1.021765 ± 0.000741 |
| 14 | 1.863426 ± 0.009652 | 1.045095 ± 0.001067 |
| 15 | 1.993101 ± 0.011115 | 1.048075 ± 0.001204 |
| 16 | 1.923772 ± 0.009008 | 1.027949 ± 0.001066 |
| 17 | 2.293430 ± 0.010012 | 1.034607 ± 0.001145 |
| 18 | 2.465899 ± 0.013529 | 1.029338 ± 0.001235 |
| 19 | 3.081268 ± 0.011685 | 1.011917 ± 0.000887 |
| 20 | 2.722161 ± 0.012073 | 1.014725 ± 0.000988 |
| 21 | 2.955917 ± 0.011301 | 1.030772 ± 0.000879 |
| 22 | 2.952362 ± 0.013987 | 1.012707 ± 0.001365 |
| 23 | 3.018248 ± 0.009408 | 1.006551 ± 0.000837 |
| 24 | 3.623165 ± 0.013787 | 1.009968 ± 0.000955 |

### 3.2 Remarks on the Memory Parameter $\mu$

From Table 1, the plot of the numerical values of $\mu$ from SC 7 to SC 24 can be viewed as a time series. As shown in Fig. 7, the memory parameter $\mu$ appears to progressively approach the value $\mu = 1$. Note that for $\mu = 1$, the sequence of sunspot numbers loses its memory, i.e., the factor, $(T - t)^{(\mu-1)/2}$, vanishes in Eq. (1). In this case, the daily sunspot numbers would behave like a memoryless Brownian motion modulated by, $e^{-\beta/2t}/t$.

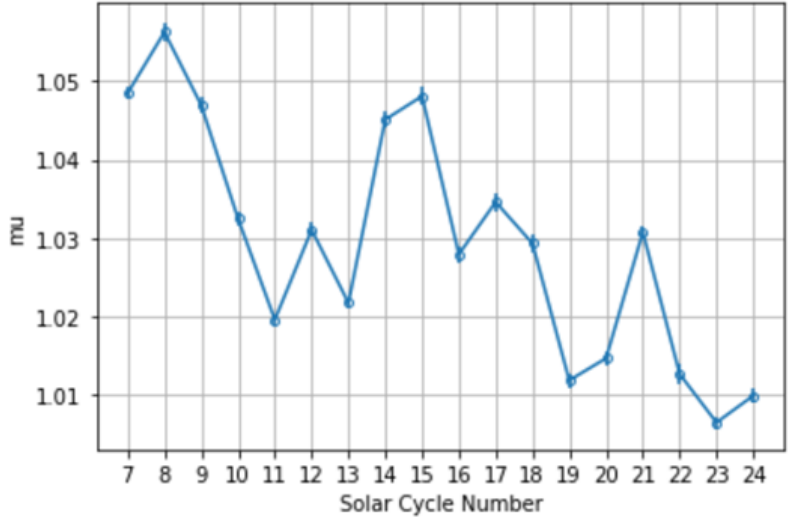

**Figure 7.** Time series of the memory parameter $\mu$

It may also be instructive to look at the theoretical MSD for large times. Taking the ln of both sides of Eq. (3) we have,

$$\ln(\text{MSD}) = \ln[a\,\Gamma(\mu)\beta^{-\mu}] + (\mu-1)\ln(T) - (\beta/T)\,. \quad (5)$$

For large time, $T \gg 1$, we can ignore the last term and obtain,

$$\ln(\text{MSD}) \approx \ln[a\,\Gamma(\mu)\beta^{-\mu}] + (\mu-1)\ln(T)\,. \quad (6)$$

A plot of Eq. (6) for $\ln(\text{MSD})$ versus $\ln(T)$ is a straight line with a slope given by, $m = (\mu-1)$. This is quite similar to fractional Brownian motion (fBm) of the form,

$\ln(\text{MSD}) \approx \alpha \ln(T)$, where a plot of $\ln(\text{MSD})$ versus $\ln(T)$ is a straight line with a slope given by $\alpha$ ($0 < \alpha < 2$). Just like fBm, the slope of the line from Eq. (6) means that for $m > 1$ one has a superdiffusive fluctuation where consecutive sunspot numbers are positively correlated. On the other hand, a slope of $m < 1$ is subdiffusive such that neighboring sunspot numbers are negatively correlated (see Fig. 8). Given the values of the memory parameter $\mu$ shown in Table 1, the slope of the line described by Eq. (6) is, $m = (\mu-1) < 1$. This means that for large times, sunspot numbers are negatively correlated exhibiting a subdiffusive behavior.

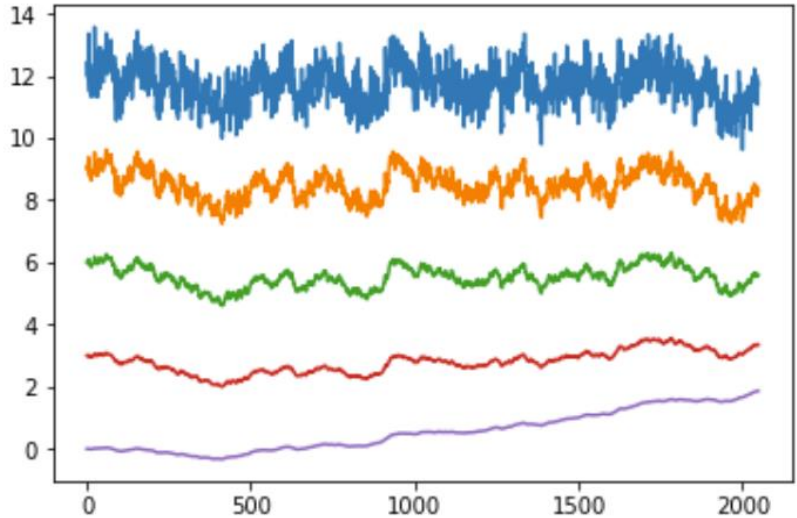

**Figure 8.** Subdiffusive fluctuation in fractional Brownian motion is shown by the upper two paths. The two lower paths exhibit superdiffusive behavior. The middle path is a memoryless ordinary Brownian motion.

### 3.3 Remarks on the Decay Parameter $\boldsymbol{\beta}$

The plot of the $\beta$ values from one solar cycle to the next is shown in Fig. 9. The plot shows a positive trend towards increasing values of $\beta$. An increase in $\beta$ value leads to a decrease in the value of the MSD as seen in Eq. (3). If this increase in the value of $\beta$ persists, this means that sunspot numbers deviate much less from a mean for each solar cycle implying a state of diminishing magnetic field turbulence or solar activity.

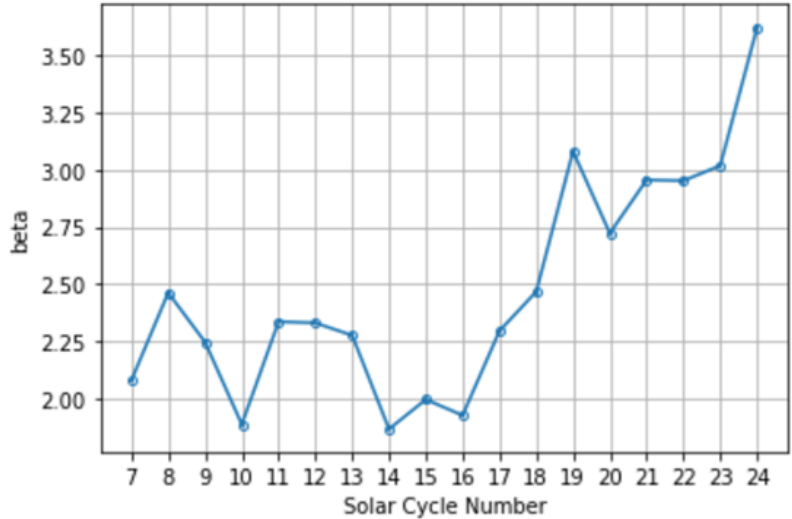

**Figure 9.** Values of $\beta$ from solar cycle 7 to solar cycle 24.

### 3.4 Probability Density Function

From the dataset, we consider again $x_j$ which is the number of sunspots at time $t_j$ in a solar cycle. We look at the difference, or displacement, $\Delta x = x_{j+\Delta} - x_j$, for a fixed lag time $\Delta$, and count the number of similar displacements as we go through different values of time $j$. We can then create a displacement distribution showing the frequency of appearance of similar values for the chosen $\Delta$ for a given solar cycle. The theoretical PDF can then be tested against the empirical displacement distribution by defining in the exponential of Eq. (2) the displacement, $\Delta x = x_f - x_i$. Using the values of the $\beta$ and $\mu$ for the solar cycle, the plot of the theoretical PDF versus $\Delta x$ can then be compared with the empirical. An example of this PDF matching is shown in Fig. 10 for solar cycle 20. Similar results are obtained for other solar cycles (see Appendix C).

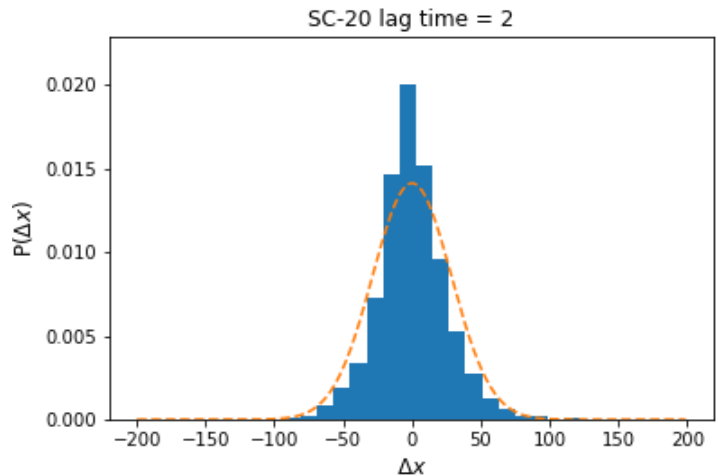

**Figure 10.** Comparing the theoretical PDF (dashed orange line), Eq. (2), with the empirical displacement distribution for solar cycle 20 for lag time equal to 2 days.

## 3.5 PDF as Function of Time

The theoretical PDF can be used to probe the behavior of sunspot numbers within a solar cycle. For instance, the daily sunspot numbers for each solar cycle can be equally divided into three segments. We label as segment 1 the first phase of a solar cycle, then segment 2 as the mid-cycle, and segment 3 the last phase of a cycle. With best-fit values for $\beta$, $\mu$, and $a$, for Eq. (3) for each segment, we can plot the theoretical PDF, Eq. (2), versus time for different values of $\Delta x = \left(x_f - x_i\right)$. This procedure enables us to characterize the phases of a solar cycle. As an example, Fig. 11 shows solar cycle 15.

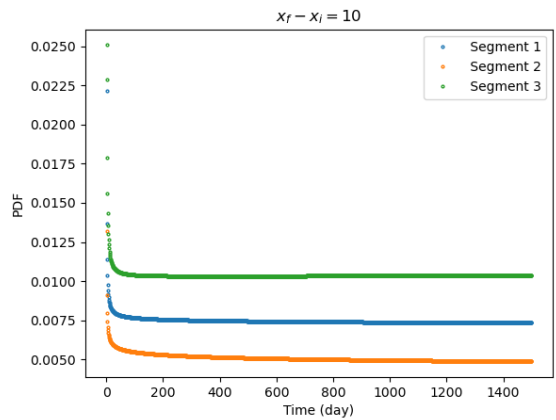

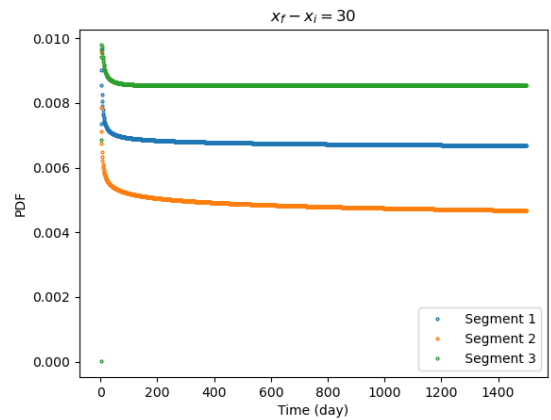

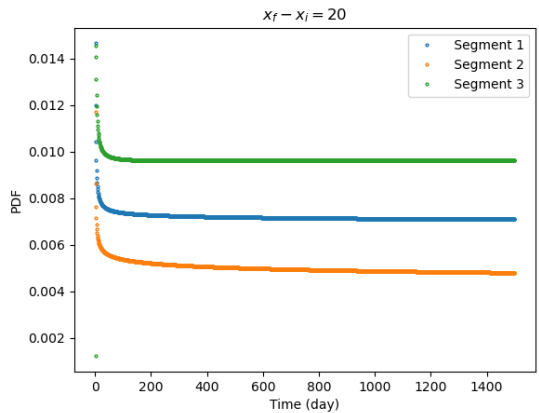

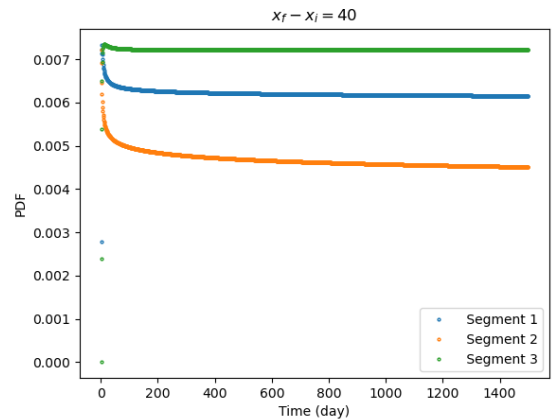

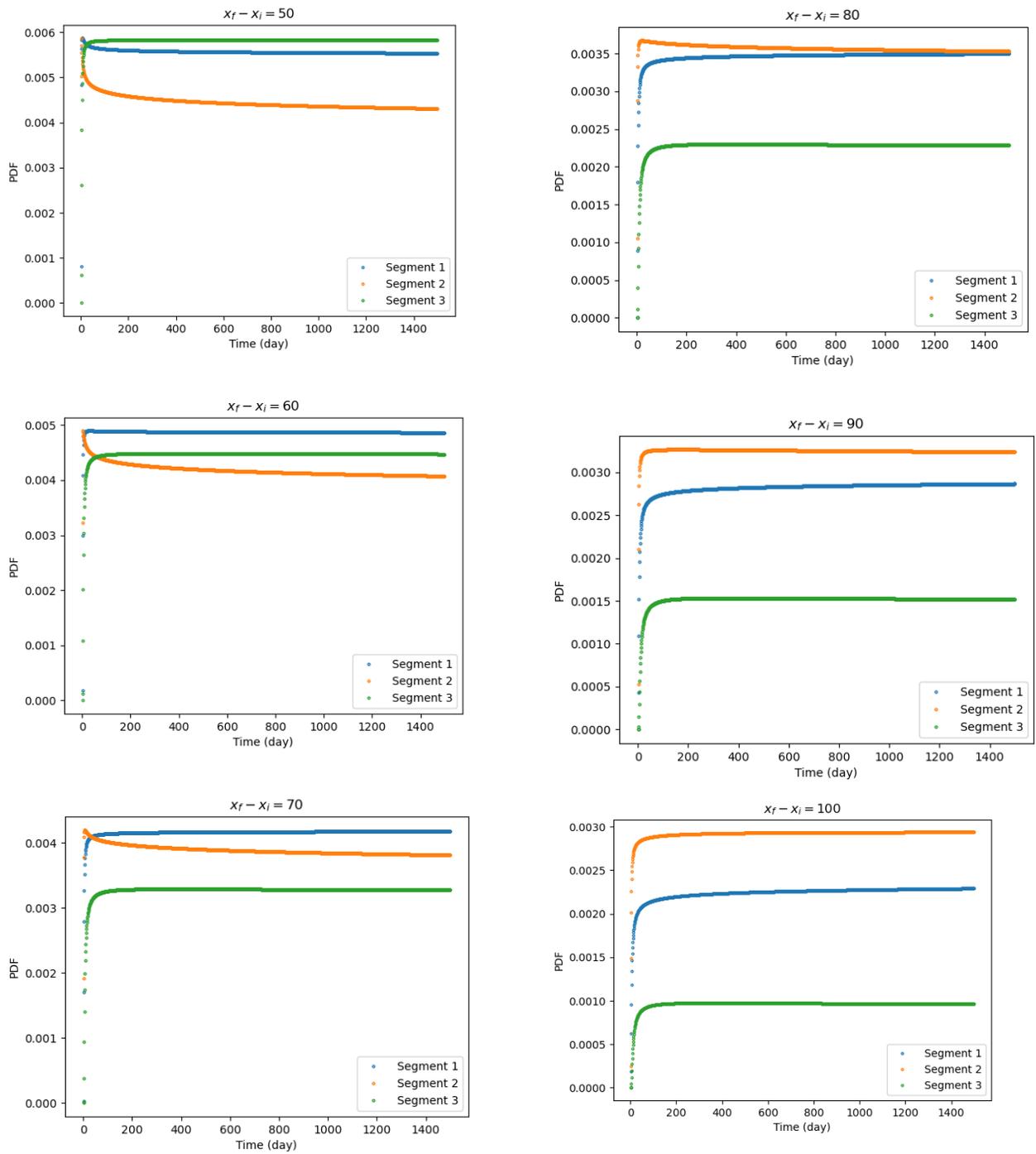

**Figure 11.** The time evolution of the PDF exhibits the behavior of sunspot numbers for solar cycle 15 for different segments or phases of the cycle.

Fig. 11 for SC 15 shows that the middle phase of the solar cycle (orange curve) is characterized by large differences in sunspot numbers, $(x_f - x_i)$, and minimal occurrence of small values of $(x_f - x_i)$. This mid-cycle is known to be the period when the north and south poles of the sun flip which manifests here as a phase with the highest probability for differences in sunspot numbers greater than, $(x_f - x_i) \approx 80$. Moreover, compared to the initial and mid-cycle phases, the last phase of a solar cycle exhibits mostly smaller differences in sunspot numbers, i.e., roughly less than $(x_f - x_i) \approx 50$. Segment 3, or the last phase of the cycle (green curve), having the highest value for $\beta$ (see, Fig. 12 and Table 2), would also consist of sunspot numbers which appear to experience the least deviation from a mean value implying a diminished solar activity. Note that, from Eq. (3), higher values of $\beta$ lead to lesser values of the MSD.

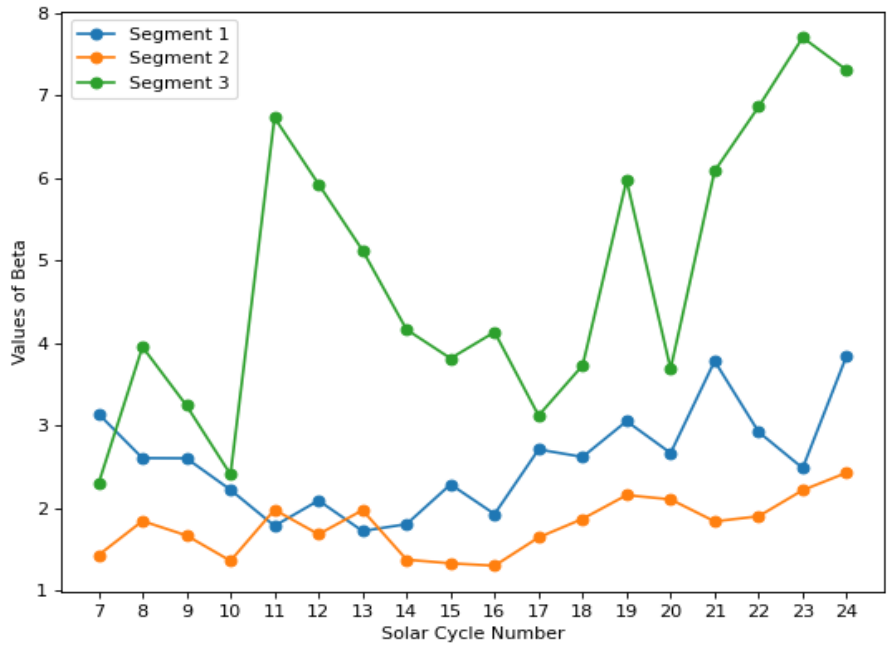

**Figure 12.** Plot of $\beta$ values for different solar cycles. Blue: segment 1; Orange: segment 2; and Green: segment 3.

**Table 2.** Values of the decay parameter $\beta$ for the three phases, or segments, of each solar cycle.

| SOLAR CYCLE | $\beta$ (SEGMENT 1) | $\beta$ (SEGMENT 2) | $\beta$ (SEGMENT 3) |
|---|---|---|---|
| 7 | 3.143018 | 1.429791 | 2.294773 |
| 8 | 2.606147 | 1.843879 | 3.954128 |
| 9 | 2.603780 | 1.668639 | 3.243757 |
| 10 | 2.222249 | 1.361203 | 2.408582 |
| 11 | 1.784081 | 1.980099 | 6.732740 |

| 12 | 2.092513 | 1.683383 | 5.927913 |
|---|---|---|---|
| 13 | 1.722498 | 1.975251 | 5.124706 |
| 14 | 1.806035 | 1.375340 | 4.164255 |
| 15 | 2.287486 | 1.331050 | 3.815064 |
| 16 | 1.928100 | 1.301723 | 4.133848 |
| 17 | 2.711008 | 1.644606 | 3.122400 |
| 18 | 2.620715 | 1.869889 | 3.729730 |
| 19 | 3.053854 | 2.157729 | 5.971996 |
| 20 | 2.667660 | 2.105974 | 3.689785 |
| 21 | 3.782743 | 1.838960 | 6.086783 |
| 22 | 2.923287 | 1.901324 | 6.865656 |
| 23 | 2.485488 | 2.217387 | 7.704668 |
| 24 | 3.848617 | 2.429626 | 7.309450 |

## 4. Conclusion

An evaluation of the mean square deviation for the fluctuating sunspot numbers from solar cycle 7 to 24 yields a novel probability density function that captures the memory behavior of the rise and fall of sunspot numbers. This was done by matching the MSD of the dataset with a theoretical MSD with a memory parameter $\mu$ and decay parameter $\beta$. The values of the $\mu$ and $\beta$ parameters for all solar cycles, when plotted as a time series, provide additional insights. The plot for $\mu$ (Fig. 7) shows a progression towards a state of diminishing memory for the fluctuation of sunspot numbers. Moreover, the values of $\mu$ indicate that, for large times, the daily sunspot numbers are negatively correlated. On the other hand, a plot of $\beta$ (Fig. 9) from one solar cycle to the next seems to indicate a diminishing solar activity. This is deduced from the theoretical MSD where increasing values of $\beta$ can lead to lesser deviations from a mean value of the sunspot numbers. The theoretical PDF is also tested against the displacement probability distribution of sunspot numbers for all solar cycles (see, e.g., Appendix C). With the PDF, a characterization of sunspot numbers for different phases of a solar cycle is done (Fig. 11) which agrees with increased solar activities accompanying the flipping of the sun's north and south poles midway through a solar cycle. It also appears that, compared with the initial and middle stages, the last phase of a solar cycle shows lesser solar activity as indicated by sunspot numbers deviating less from a mean value. This novel PDF provides a new perspective that can be utilized in further investigations for a better understanding of solar dynamics.

**Conflict of Interest**

The authors declare that the research was conducted in the absence of any commercial or financial relationships that could be construed as a potential conflict of interest.

**Author Contributions**

R.T. conceptualized, curated the data, generated the plots, and reviewed the paper. R.R. conceptualized, generated graph, supervised, and reviewed the paper. C. C. B. conceptualized, supervised, wrote, and reviewed the paper.

## ACKNOWLEDGEMENTS

Helpful discussions with Georgiy Shevchenko are gratefully acknowledged. R.L.T. wishes to thank the Department of Science and Technology (DOST ASTHRDP-NSC) for its support.

## Data Availability Statement

The datasets generated and analyzed for this study can be found in the World Data Center SILSO, Royal Observatory of Belgium, Brussels accessible at this link: https://www.sidc.be/silso/datafiles

## REFERENCES

Arlt R., Leussu R., Giese N., Mursula K. and Usoskin I. G. (2013). Sunspot positions and sizes for 1825- 1867 from the observations by Samuel Heinrich Schwabe. Mon. Not. R. Astron. Soc., 433, 3165–3172. [arXiv:1305.7400 [astro-ph.SR]].

Aschwanden and Johnson (2021). The Solar Memory from Hours to Decades. Astrophys. Jour., 921:82. https://doi.org/10.3847/1538-4357/ac2a29

Aure R. R. L., Bernido C. C., Carpio-Bernido M. V., Bacabac R. G. (2019). Damped white noise diffusion with memory for diffusing microprobes in ageing fibrin gels. Biophys. Jour., 117, 1029–1036. https://doi.org/10.1016/j.bpj.2019.08.014

Barredo W., Bernido C.C., Carpio-Bernido M.V., and Bornales J.B. (2018). Modelling non-Markovian fluctuations in intracellular biomolecular transport. Math.Biosci. 297, 27-31

Berdyugina S. V. and Usoskin I. G. (2003). Active longitudes in sunspot activity: Century scale Persistence. Astron. Astrophys., 405, 1121–1128.

Bernido C.C. and Carpio-Bernido M.V. (2015). Methods and Applications of White Noise Analysis in Interdisciplinary Sciences. Singapore: World Scientific.

Bernido C.C. and Carpio-Bernido M.V. (2012). White noise analysis: some applications in complex systems, biophysics and quantum mechanics. Int. J. Mod. Phys. B, 26(12300014).

Bushby P. and Mason J. (2004). Understanding the solar dynamo. Astron. & Geophys. 45, 4.7–4.13. https://doi.org/10.1046/j.1468-4004.2003.45407.x

Bernido C.C., Carpio-Bernido M.V., and Escobido M.G.O. (2014). Modified diffusion memory for cyclone track fluctuations. Phys. Lett. A, 378, 2016-2019. https://doi.org/10.1016/j.physleta.2014.06.003

Choudhuri A.R., Karak B.B. Origin of grand minima in sunspot cycles. Phys Rev Lett. 2012 Oct 26;109(17):171103. doi: 10.1103/PhysRevLett.109.171103. Epub 2012 Oct 25. PMID: 23215173.

Elnar A.R.B., Cena C.B., Bernido C.C., and Carpio-Bernido M.V. (2021). Great Barrier Reef degradation, sea surface temperatures, and atmospheric CO2 levels collectively exhibit a stochastic process with memory. Climate Dynamics, 57, 2701-2711. https://doi.org/10.1007/s00382-021-05831-8

Goelzer ML et al (2013) An analysis of heliospheric magnetic field flux based on sunspot number from 1749 to today and prediction for the coming solar minimum. J Geophys Res 118: 7525-7531

Hathaway D.H. (2015). The Solar Cycle. Living Rev. Solar Phys., 12, 4. DOI 10.1007/lrsp-2015-4

Hida T., Kuo H.H., Potthoff J., and Streit L. (2013). White noise: an infinite dimensional calculus (Vol. 253). Springer Science & Business Media.

Karak BB and Nandy D. (2012). Astrophys. Jour. Lett, 761:L13 . doi:10.1088/2041-8205/761/1/L13

Liu Y.D. et al (2014). Observations of an extreme storm in interplanetary space caused by successive coronal mass ejections. Nat. Comm. 5:3481. DOI: 10.1038/ncomms4481

Mohaved M.S., Jafari G.R., Ghasemi F., Rahvar S., Tabar M.R.R. (2006). Multifractal detrended fluctuation analysis of sunspot time series, J. Stat. Mech.: Theory Exp. 2006. P02003. DOI: 10.1088/1742-5468/2006/02/P02003.

Muñoz-Jaramillo A., et al. (2013). Solar Cycle Propagation, Memory, and Prediction: Insights from a Century of Magnetic Proxies. Astrophys. Jour. Lett. 767:L25. doi:10.1088/2041-8205/767/2/L25

Okamoto T.J., and Sakura, T. (2018). Super-strong Magnetic Field in Sunspots. The Astrophysical Journal Letters, 852:L16. DOI 10.3847/2041-8213/aaa3d8

Reyes R. and Bernido C.C. (2023). A Model-Independent Determination of Red Noise in Pulsar Timing Arrivals. http://arxiv.org/abs/2303.00931

Richardson I. G. and Cane H. V. (2010). Near-Earth interplanetary coronal mass ejections during solar cycle 23 (1996–2009): Catalog and summary of properties. Solar Phys. 264, 189–237.

Smith E.J. and Balogh A. (2008), Decrease in heliospheric magnetic flux in this solar minimum: Recent Ulysses magnetic field observations, Geophys. Res. Lett., 35, L22103, doi:10.1029/2008GL035345.

Violanda R.R., Bernido C.C., and Carpio-Bernido M.V. (2019). White noise functional integral for exponentially decaying memory: nucleotide distribution in bacterial genomes. Physica Scripta, 94, 125006. https://doi.org/10.1088/1402-4896/ab3739

Yeates A.R., Nandy D., and Mackay D.H. (2008). Exploring The Physical Basis Of Solar Cycle Predictions: Flux Transport Dynamics And Persistence Of Memory In Advection- Versus Diffusion-Dominated Solar Convection Zones. Astrophys. Jour. 673, 544-556.

## APPENDIX A: Probability Density Function

As parametrized by Eq. (1), the fluctuating sunspot numbers represented by variable $x$ starting at $x_i$ can end anywhere at some final time $T$. Using the Donsker delta function (Hida et al 1993), we can pin down the final point such that at time $T$ it ends at, $x(T) = x_f$, i.e.,

$$\delta\big(x(T) - x_f\big) = \delta\left(x_i + \sqrt{a}\int_0^T (T-t)^{(\mu-1)/2}\right.$$

$$\left. \times \frac{exp(-\beta/2t)}{t^{(\mu+1)/2}} dB(t) - x_f\right)$$

(A1)

In Eq. (A1), we have written $x(T)$ explicitly using Eq. (1). For all fluctuating paths that do not end at $x_f$, the delta function Eq. (A1) vanishes. Note that Eq. (A1) can be written as a functional of the white noise random variable $\omega(t)$ (Hida et al 1993) where, $\omega(t) = dB(t)/dt$. One can then obtain the probability density function, $P(x_f, T; x_i, 0)$, that a path ends at $x_f$ if it started at $x_i$, by integrating over all possible paths that satisfy Eq. (A1), i.e.,

$$P(x_f, T; x_i, 0) = \int \delta(x(T) - x_f) \, d\mu(\omega) \,, \tag{A2}$$

where $d\mu(\omega)$ is the Gaussian white noise measure (Hida et al 1993). Expressing the delta function in terms of its Fourier representation we have,

$$P(x_f, T; x_i, 0)$$
$$= \frac{1}{\sqrt{2\pi}} \int \int_{-\infty}^{+\infty} e^{ik(x(T) - x_f)} dk \, d\mu(\omega)$$
$$= \frac{1}{\sqrt{2\pi}} \int_{-\infty}^{+\infty} dk \, e^{ik(x_i - x_f)} \times$$

$$\int e^{ik\left[\sqrt{a} \int_0^T (T-t)^{(\mu-1)/2} \frac{exp(-\beta/2\tau)}{t^{(\mu+1)/2}} \omega(t) dt\right]} d\mu(\omega)$$
$$= \frac{1}{\sqrt{2\pi}} \int_{-\infty}^{+\infty} dk \, e^{ik(x_i - x_f)}$$

$$\times \int e^{i \int_0^T \omega(t) \, \xi(t) dt} \, d\mu(\omega) \,, \tag{A3}$$

where, $\xi(t) = k\sqrt{a}(T-t)^{(\mu-1)/2} exp(-\beta/2t)/t^{(\mu+1)/2}$. Integration over $d\mu(\omega)$ is done using the characteristic functional (Hida et al 1993),

$$\int e^{i \int_0^T \omega(t) \, \xi(t) dt} \, d\mu(\omega) = e^{-\frac{1}{2} \int_0^T \xi(t)^2 d\tau} \,, \tag{A4}$$

to get,

$$P(x_f, T; x_i, 0) = \frac{1}{\sqrt{2\pi}} \int_{-\infty}^{+\infty} e^{ik(x_i - x_f)}$$
$$\times e^{-\frac{k^2}{2}\left[a \int_0^T (T-t)^{\mu-1} exp(-\beta/t)/t^{\mu+1} dt\right]} dk \tag{A5}$$

The integral over $dk$ is a Gaussian integral which yields $P(x_f, T; x_i, 0)$ given by Eq. (2).

## APPENDIX B: Theoretical and Empirical MSD for Solar Cycles 7 to 24

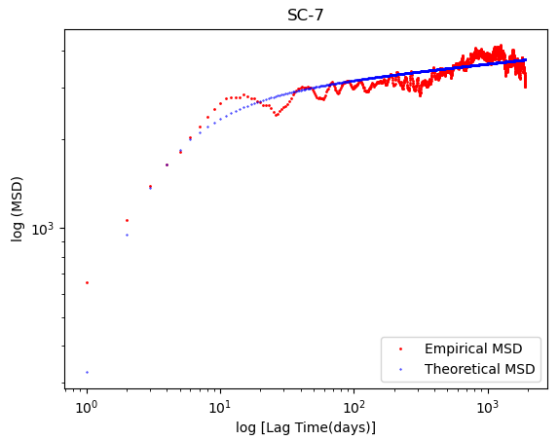

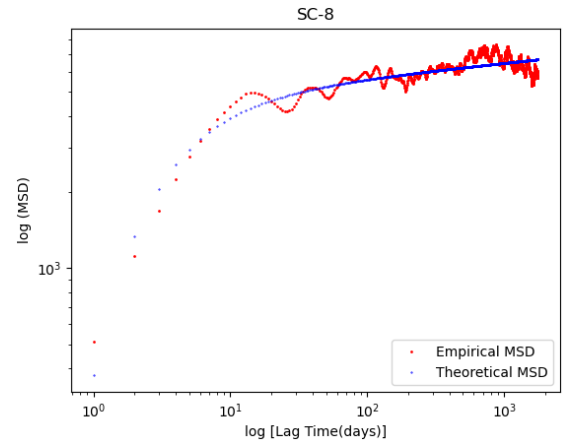

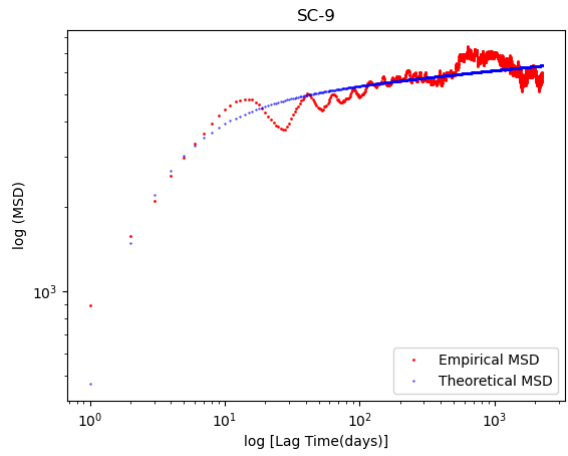

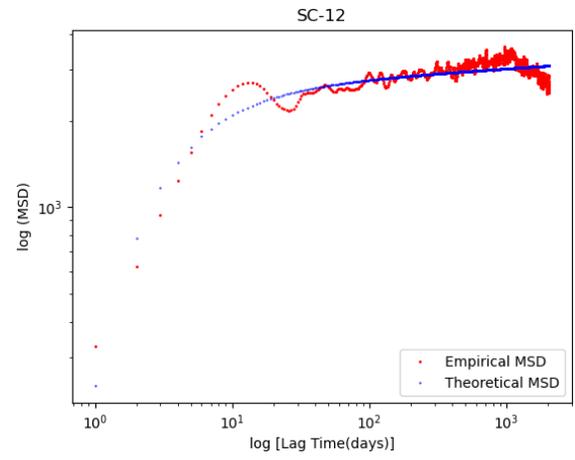

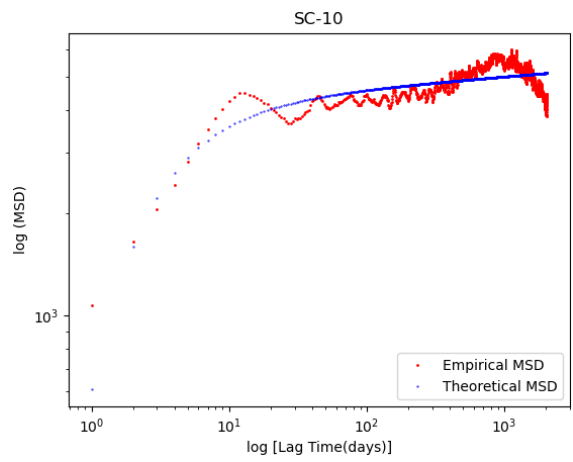

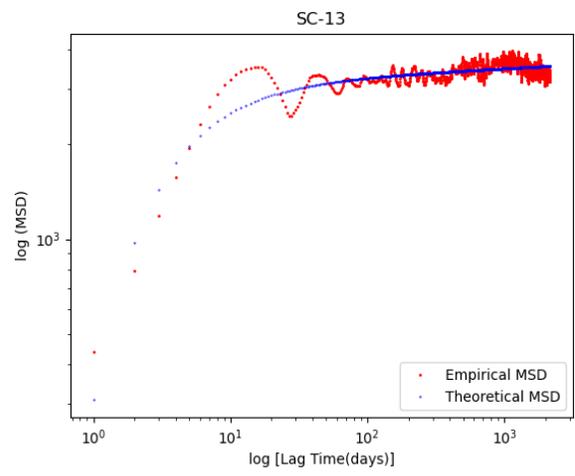

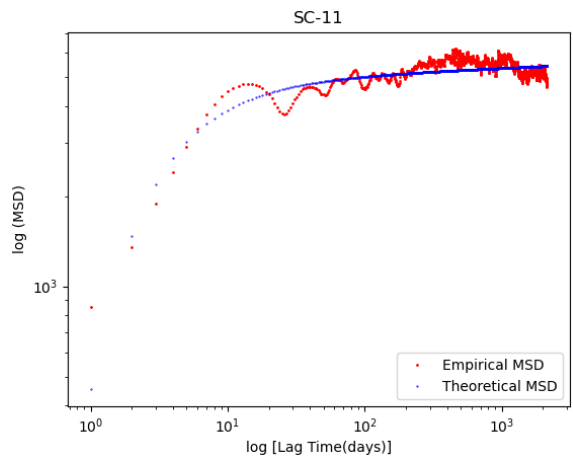

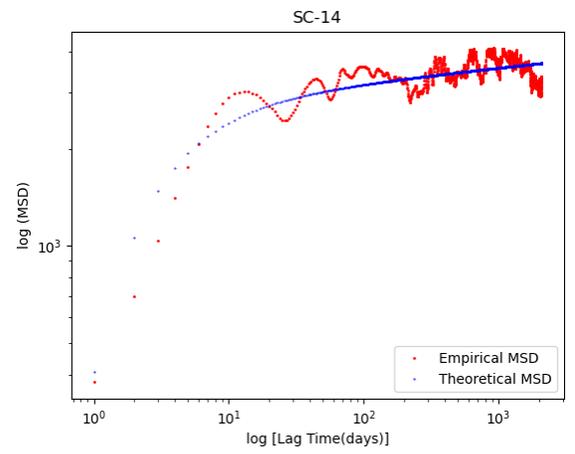

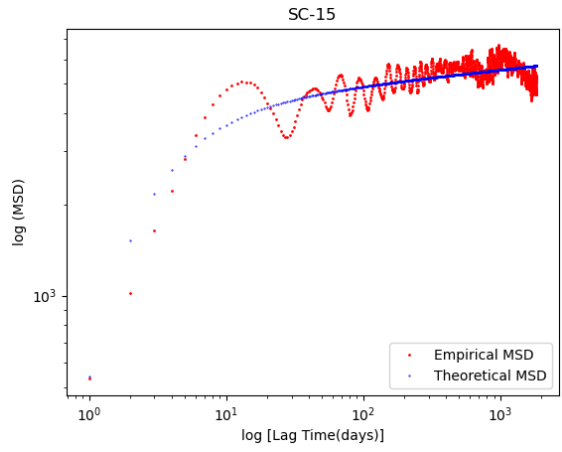

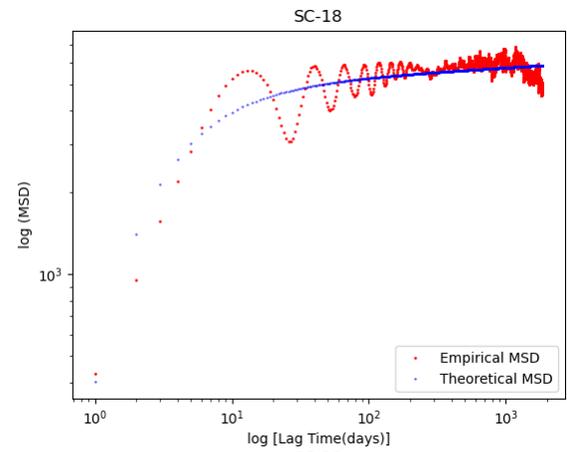

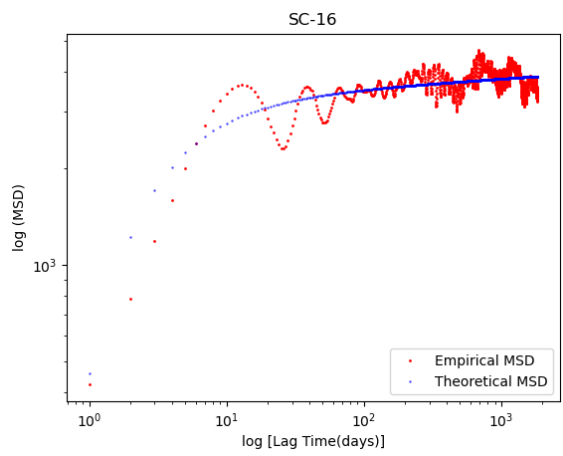

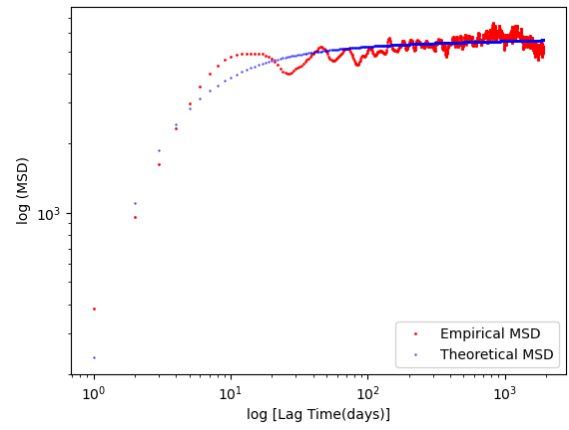

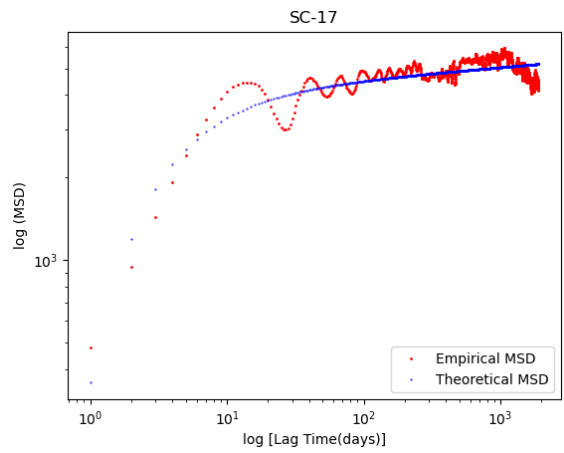

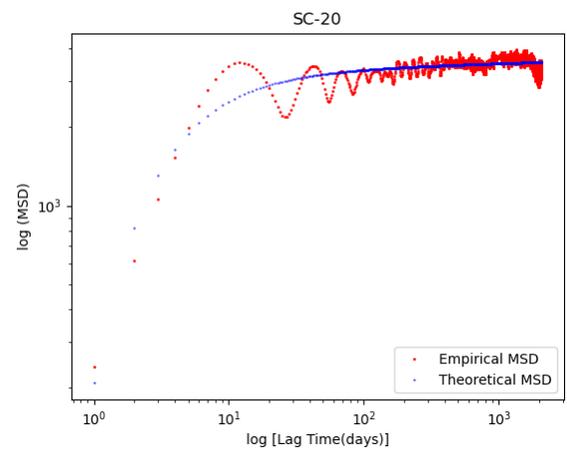

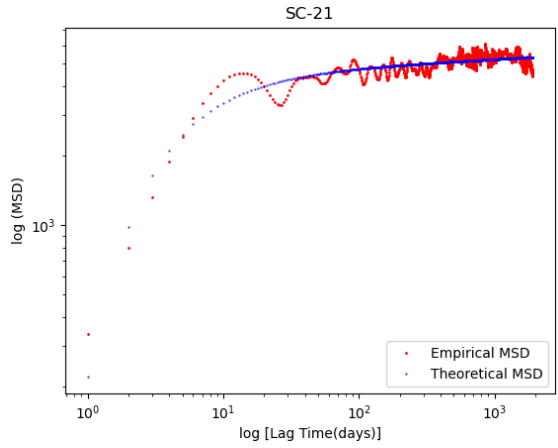

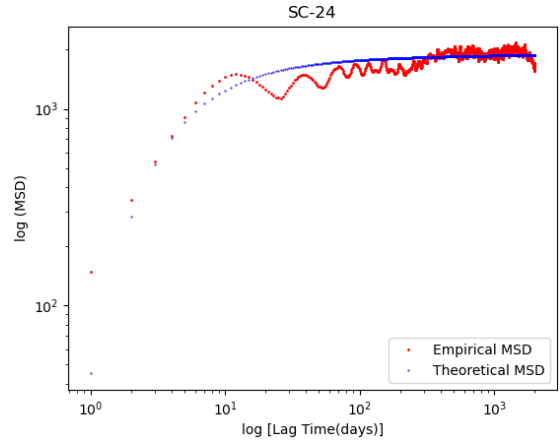

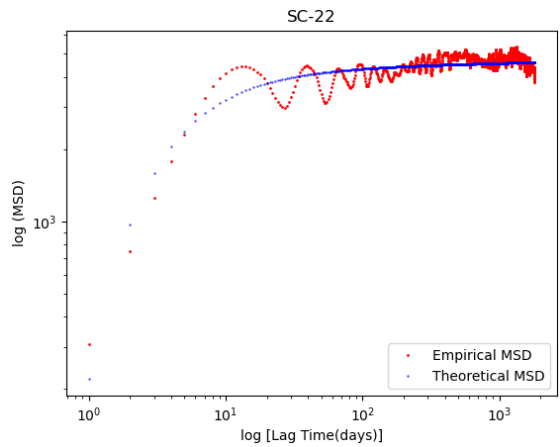

## APPENDIX C: Theoretical and Empirical PDF for Solar Cycles 7 to 24 for Lag Time = 2

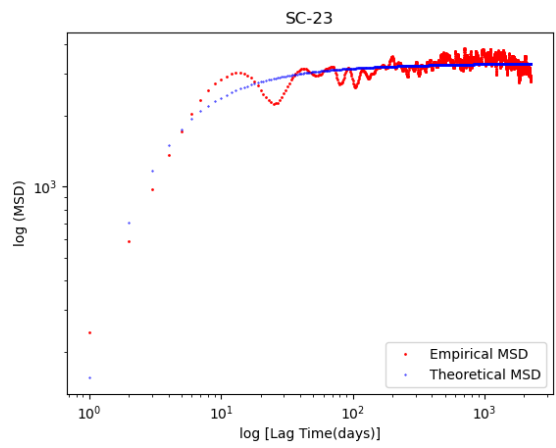

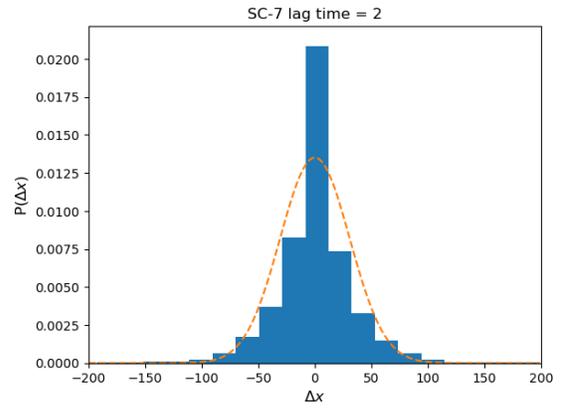

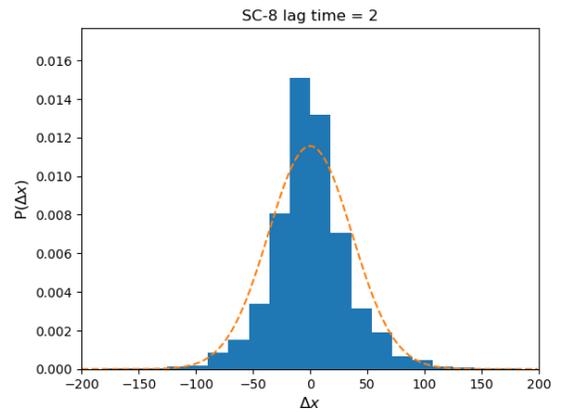

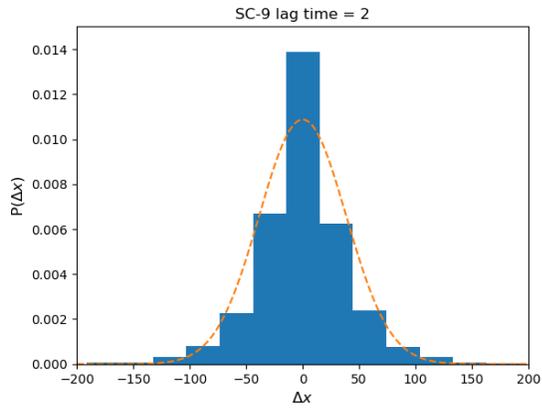

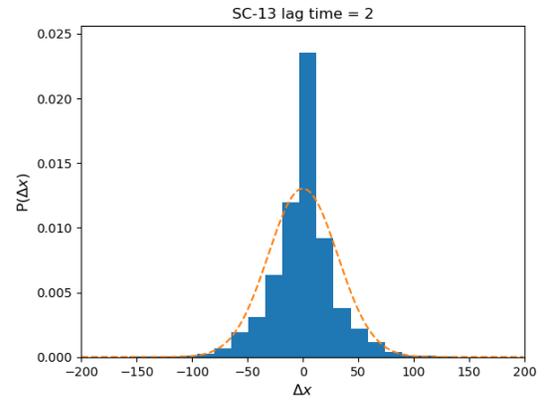

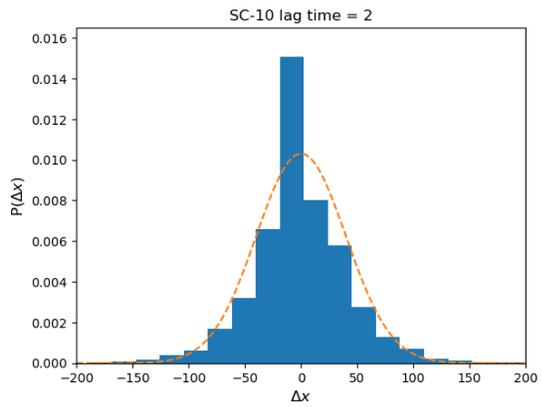

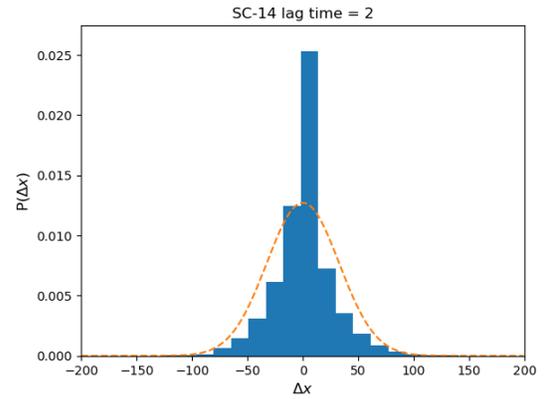

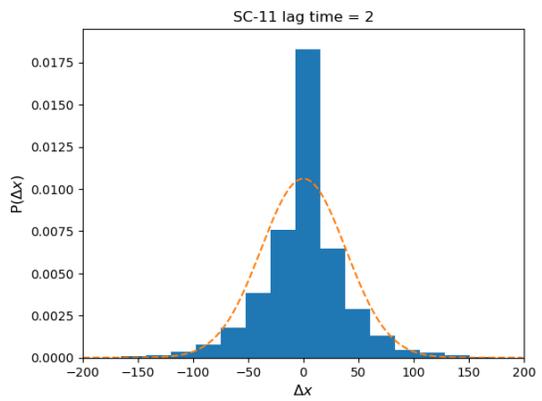

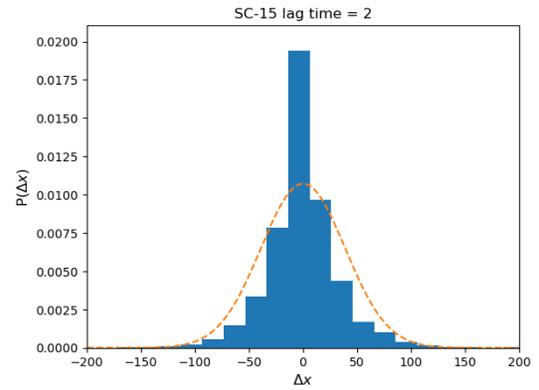

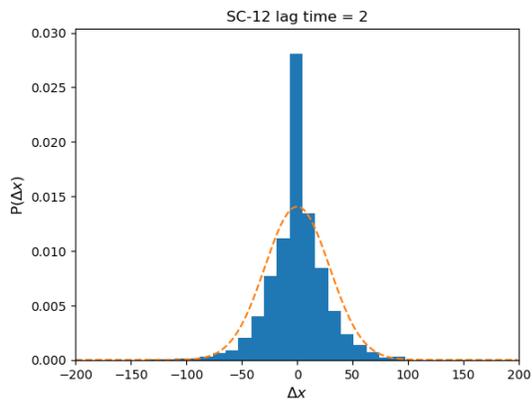

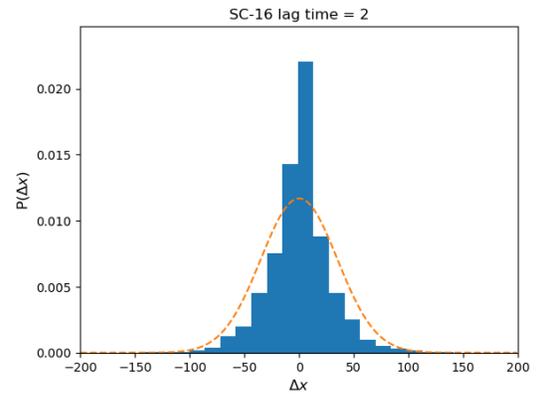

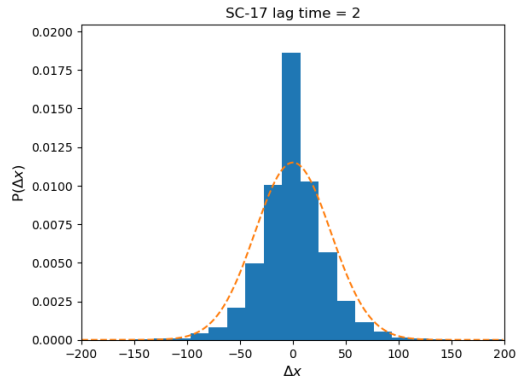
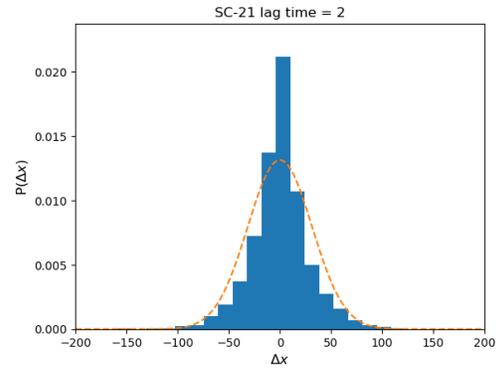
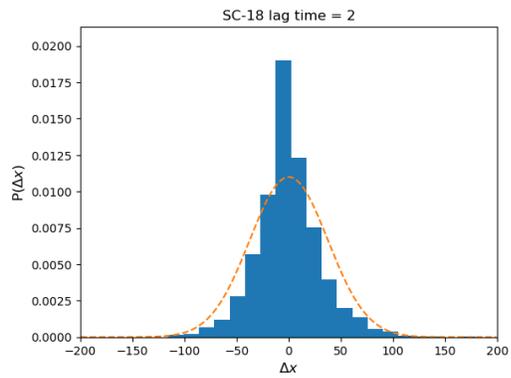
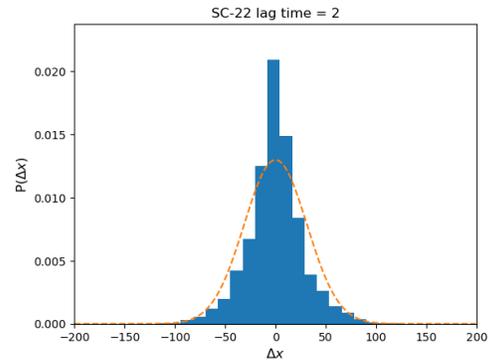
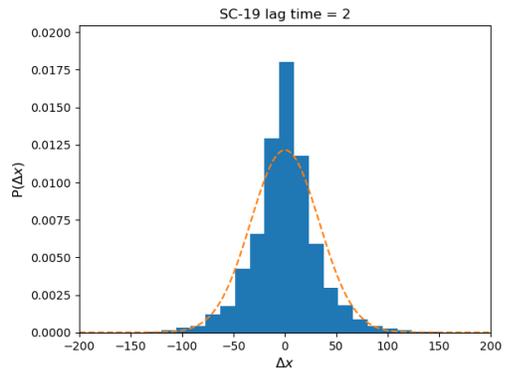
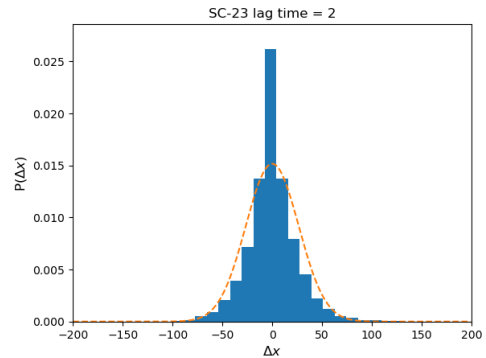
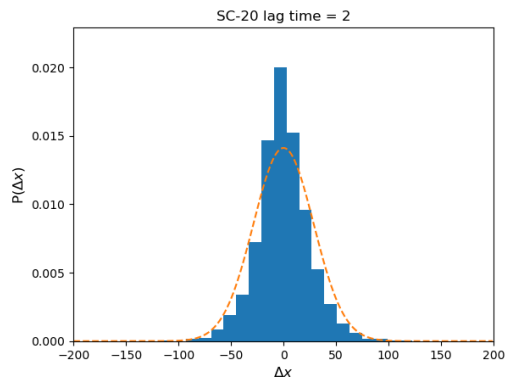
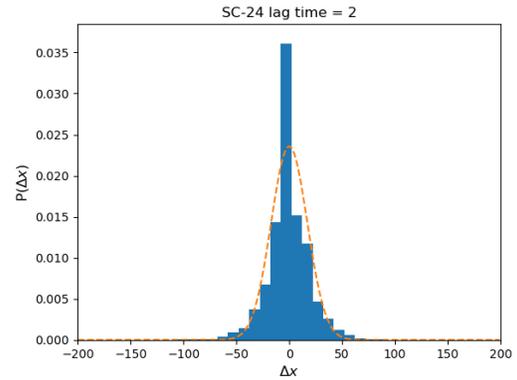